\documentclass[unnumsec,webpdf,contemporary,large]{oup-authoring-template}%

\graphicspath{{Fig/}}

\theoremstyle{thmstyleone}%
\theoremstyle{thmstyletwo}%
\theoremstyle{thmstylethree}%

\usepackage[utf8]{inputenc} 
\usepackage{amsfonts}       
\usepackage{nicefrac}       
\usepackage{microtype}      
\usepackage{xcolor}         
\usepackage{amsmath}
\usepackage{comment}
\usepackage{graphicx}
\usepackage{multirow}
\usepackage[normalem]{ulem}
\useunder{\uline}{\ul}{}

\begin{document}

\journaltitle{Journal Title Here}
\DOI{DOI HERE}
\copyrightyear{2022}
\pubyear{2019}
\access{Advance Access Publication Date: Day Month Year}
\appnotes{Paper}

\firstpage{1}


\title[BoKDiff]{BoKDiff: Best-of-K Diffusion Alignment for Target-Specific 3D Molecule Generation}

\author[1]{Ali Khodabandeh Yalabadi}
\author[2]{Mehdi Yazdani-Jahromi}
\author[1]{Ozlem Ozmen Garibay}

\authormark{A.K. Yalabadi et al.}

\address[1]{\orgdiv{Department of Industrial Engineering and Management Systems}, \orgname{University of Central Florida}, \orgaddress{\street{4000 Central Florida Blvd}, \postcode{32816}, \state{FL}, \country{USA}}}
\address[2]{\orgdiv{Department of Computer Science}, \orgname{University of Central Florida}, \orgaddress{\street{4000 Central Florida Blvd}, \postcode{32816}, \state{FL}, \country{USA}}}

\corresp[$\ast$]{Corresponding author. \href{email:email-id.com}{ozlem@ucf.edu}}

\received{Date}{0}{Year}
\revised{Date}{0}{Year}
\accepted{Date}{0}{Year}



\abstract{
Structure-based drug design (SBDD) leverages the 3D structure of biomolecular targets to guide the creation of new therapeutic agents. Recent advances in generative models, including diffusion models and geometric deep learning, have demonstrated promise in optimizing ligand generation. However, the scarcity of high-quality protein-ligand complex data and the inherent challenges in aligning generated ligands with target proteins limit the effectiveness of these methods. We propose BoKDiff, a novel framework that enhances ligand generation by combining multi-objective optimization and Best-of-K alignment methodologies. Built upon the DecompDiff model, BoKDiff generates diverse candidates and ranks them using a weighted evaluation of molecular properties such as QED, SA, and docking scores. To address alignment challenges, we introduce a method that relocates the center of mass of generated ligands to their docking poses, enabling accurate sub-component extraction. Additionally, we integrate a Best-of-N (BoN) sampling approach, which selects the optimal ligand from multiple generated candidates without requiring fine-tuning. BoN achieves exceptional results, with QED values exceeding 0.6, SA scores above 0.75, and a success rate surpassing 35\%, demonstrating its efficiency and practicality. BoKDiff achieves state-of-the-art results on the CrossDocked2020 dataset, including a -8.58 average Vina docking score and a 26\% success rate in molecule generation. This study is the first to apply Best-of-K alignment and Best-of-N sampling to SBDD, highlighting their potential to bridge generative modeling with practical drug discovery requirements. The code is provided at this \href{https://github.com/khodabandeh-ali/BoKDiff.git}{GitHub URL}.}


\maketitle

\section{Introduction}
Structure-based drug design (SBDD) \cite{anderson2003process} leverages the 3D structures of biomolecules to facilitate the development and refinement of therapeutic agents. The primary aim is to design molecules that specifically interact with target proteins. Recent advancements have reframed this as a data-driven, conditional generative task, employing geometric deep learning to enhance molecular generation \cite{powers2023geometric}. Strategies like SE(3)-equivariant auto-regressive models for sequentially generating atomic or fragment structures \cite{peng2022pocket2mol, zhang2023learning} and diffusion models for predicting atomic types and positions \cite{luo20213d, guan20233d} have shown promise.

Despite these innovations, the limited availability of high-quality protein-ligand complex data remains a key challenge \cite{vamathevan2019applications}. Deep learning’s effectiveness often hinges on large-scale datasets, a luxury readily available in domains like computer vision and natural language processing due to accessible data sources. In contrast, generating protein-ligand binding data is inherently complex and resource-intensive. Datasets such as CrossDocked2020 \cite{francoeur2020three} address this scarcity by augmenting data through docking ligands into diverse binding pockets. While this expands dataset size, it introduces potential compromises in data quality, as ligands may exhibit only moderate binding affinities, insufficient for rigorous drug design demands \cite{zhou2024decompopt}. Additionally, this method fails to enhance the diversity of unique ligands, restricting generative models’ ability to learn from diverse molecular structures.

To overcome these limitations, approaches like property-based molecule search \cite{fu2022reinforced} have explored vast chemical spaces for molecules with desirable traits. However, such methods lack the diversity and creativity of generative models. Conditional diffusion models integrated with iterative substructure optimization \cite{zhou2024decompopt} have demonstrated improved molecular properties while preserving some diversity. Yet, these methods are constrained by static model parameters during optimization, which may limit their overall performance.

The challenge of optimizing molecular properties remains significant. To address this, we propose BoKDiff, a multi-objective optimization framework that aligns and enhances existing diffusion models. Drawing inspiration from the effective RAFT (Reward rAnked FineTuning) alignment method \cite{dong2023raft}, our approach uses a Best-of-K methodology to curate high-reward subsets based on molecular properties. These curated subsets are then employed to fine-tune the diffusion models for generating ligands with improved molecular properties while maintaining or enhancing their docking performance.

In addition, we explore the Best-of-N (BoN) sampling approach \cite{gui2024bonbon} as a complementary strategy for optimizing ligand generation. Unlike BoKDiff, BoN does not involve fine-tuning; instead, it directly selects the highest-reward ligand from N generated candidates based on a multi-objective reward function. This method offers improved molecular property results, such as QED over 0.6, SA above 0.75, and a success rate surpassing 35\%. However, BoN’s computational cost is higher, making it less general and scalable compared to BoKDiff.

Our study adopts the \textbf{DecompDiff} model \cite{guan2024decompdiff} as the foundation, which generates candidate molecules by leveraging sub-pocket and sub-structure information, such as arms and scaffolds, from protein-ligand complexes. However, a key limitation of AI-generated ligands is their misalignment with target proteins, leading to failed sub-component extractions. To overcome this, we relocate the center of mass of generated ligands to align with the docking pose, ensuring proximity to the target protein. BoKDiff demonstrates superior performance, significantly outperforming baseline methods. Below, we summarize the key contributions of our work:

\begin{itemize}
    \item Our method surpasses the state-of-the-art DecompDiff model across multiple critical metrics, achieving a success rate increase from 24.5\% to 26\%, a QED improvement up to 13\%, and Vina-related score enhancements up to 6\%. These results highlight the effectiveness of our approach in generating higher-quality, more drug-like molecular candidates.
    \item We introduce multi-objectivity to this domain and use three metrics, QED, SA, and Vina Docking Score (see \nameref{evaluation} section) to evaluate the generated samples and rank them based on a final reward value in a weighted sum manner.
    \item We solve the proximity issue by relocating the selected best-of-K ligands to the center of mass of their docking poses, addressing alignment issues for sub-component extraction.
    \item BoKDiff achieves a -8.58 average Vina Docking score and a 26\% success rate for molecule generation on the CrossDocked2020 dataset.
    \item We explored the Best-of-N (BoN) sampling approach, achieving superior QED, SA, and success rate values, and demonstrated its complementary potential to BoKDiff.
    \item To the best of our knowledge, this is the first application of best-of-K alignment and BoN sampling to structure-based drug design, aligning generative models with the practical requirements of drug discovery.
\end{itemize}

\section{Related works}
One significant area of focus in molecular design is the development of ligand molecules that can effectively bind to specific target proteins, commonly referred to as Structure-Based Drug Design (SBDD). In recent years, numerous efforts have been made to improve the process of generating molecules with desirable properties. For instance, Ragoza et al. \cite{ragoza2022generating} utilized variational autoencoders to produce 3D molecular structures represented as atomic density grids. Other approaches, such as those by Luo et al. \cite{luo20213d}, Peng et al. \cite{peng2022pocket2mol}, and Liu et al. \cite{liu2022generating}, have employed autoregressive techniques to construct 3D molecules incrementally, adding one atom at a time. Similarly, Zhang et al. \cite{zhang2023molecule} introduced a fragment-based autoregressive method, predicting molecular components sequentially. Recently, diffusion models have gained traction in SBDD, with studies by Guan et al. \cite{guan20233d}, Schneuing et al. \cite{schneuing2022structure}, and Lin et al. \cite{lin2022diffbp} using these models to first generate atomic positions and types, followed by a post-processing step to define bond types. To further enhance SBDD methodologies, some research has incorporated biochemical knowledge as priors. For example, DecompDiff \cite{guan2024decompdiff} employs diffusion models to simultaneously generate atoms and bonds, guided by decomposed structural priors and validity constraints, while DrugGPS \cite{zhang2023learning} leverages subpocket-level similarities to enrich molecular generation by aligning subpocket prototypes with molecular motifs.

Beyond generative modeling of protein-ligand complexes, several researchers have integrated optimization techniques to tailor molecules with specific attributes. AutoGrow 4 \cite{spiegel2020autogrow4} and RGA \cite{fu2022reinforced} exemplify this by using genetic algorithms to enhance ligand binding affinities for target proteins. RGA further interprets the evolutionary process as a Markov decision process, steering it with reinforcement learning. Similarly, DecompOpt \cite{zhou2024decompopt} employs a controllable diffusion model that generates molecules based on protein subpockets and reference substructures, coupled with iterative optimization to refine desired properties by iteratively conditioning on previously generated substructures. Our research also targets the optimization of molecular properties, but rather than directly optimizing the molecules themselves, we focus on refining the parameters of the model responsible for generating them.

\paragraph{\textbf{Alignment via Human/AI Feedback}}
Generative models often fail to fully capture user preferences when optimized solely for likelihood. To address this, reinforcement learning with human or AI feedback has been used to align models with user preferences \cite{ziegler2019fine, ouyang2022training}. This involves training a reward model from comparison data and fine-tuning the model using policy-gradient methods \cite{christiano2017deep, schulman2017proximal}. Similar approaches have been applied to diffusion models in text-to-image generation, treating the process as a multi-step Markov decision task for refinement \cite{black2023training, zhang2024large}.

Recently, Direct Preference Optimization (DPO) \cite{rafailov2024direct} has emerged as a prominent method for aligning large language models with human preferences by directly optimizing on comparison data. Wallace et al. \cite{wallace2024diffusion} extended this approach to diffusion models, introducing Diffusion-DPO for aligning text-to-image generation with human preferences. While these methods predominantly target language models and text-to-image generators, Zhou et al. \cite{zhou2024antigen} applied DPO to fine-tune diffusion models for antibody design, using low Rosetta energy as the preference criterion. Furthermore, RAFT \cite{dong2023raft} proposed a reward-ranked fine-tuning approach that reduces memory and computational demands while remaining effective. This method selects high-reward samples to iteratively update the model’s parameters, progressively improving its performance.
In our work, we leverage best-of-K alignment, as introduced in RAFT \cite{dong2023raft}, to enhance the properties of molecules generated for specific protein binding pockets. Our primary focus is on optimizing molecular properties such as drug-likeness and synthesizability to meet desired criteria.

\section{Methodology}
In this section, we introduce BoKDiff, our proposed method for aligning diffusion models to optimize key molecular structure metrics, including drug-likeness (QED), synthesizability (SA), and docking scores, approximated using the Vina Docking score. The section is organized as follows: we begin by defining the Structure-Based Drug Design (SBDD) task and outlining the decomposed diffusion model in \nameref{decompdiff} section. Next, we provide an overview of the Best-of-K alignment method, which leverages reward ranking and fine-tuning, in \nameref{raft} section. Finally, we delve into the implementation details of our approach in \nameref{implementation} section.

\subsection{DecompDiff}
\label{decompdiff}
In the context of structure-based drug design (SBDD), generative models are conditioned on a protein's binding site, denoted as 
$\mathcal{P} = \{(x_i^\mathcal{P}, v_i^\mathcal{P})\}_{i \in \{1, \dots, N_\mathcal{P}\}}$, to produce ligands $\mathcal{M} = \{(x_i^\mathcal{M}, v_i^\mathcal{M}, b_{ij}^\mathcal{M})\}_{i,j \in \{1, \dots, N_\mathcal{M}\}}$ that can interact with the site. Here, $N_\mathcal{P}$ and $N_\mathcal{M}$ represent the number of atoms in the protein and ligand, respectively. The variables $x \in \mathbb{R}^3$ and $v \in \mathbb{R}^h$ correspond to the spatial coordinates of the atoms and their types, while $b_{ij} \in \mathbb{R}^5$ describes the bonds between atoms. In this work, we consider $h$ distinct atom types (e.g., H, C, N, O, S, Se) and 5 categories of bonds: non-bonded, single, double, triple, and aromatic.

Following the framework of the decomposed diffusion model proposed by Guan et al. \cite{guan2024decompdiff}, each ligand is divided into fragments $\mathcal{K}$, consisting of multiple arms $\mathcal{A}$ connected by at most one scaffold $\mathcal{S}$, where $|\mathcal{A}| \geq 1$, $|\mathcal{S}| \leq 1$, and $\mathcal{K} = |\mathcal{A}| + |\mathcal{S}|$. Leveraging these decomposed substructures, the model employs informative, data-dependent priors $\mathcal{O}_\mathcal{P} = \{\mu_{1:K}, \Sigma_{1:K}, \mathcal{H}\}$, which are derived from atom positions through maximum likelihood estimation. These priors are designed to enhance the training of the diffusion model. 

The generative process gradually diffuses the molecular structure $\mathcal{M}$ over a fixed schedule $\{\lambda_t\}_{t=1,\dots,T}$. Let $\alpha_t = 1 - \lambda_t$ and $\tilde{\alpha}_t = \prod_{s=1}^t \alpha_s$. For each atom, its position at time $t$ is shifted to a prior-centered position $\tilde{x}_t^i = x_t^i - (\mathcal{H}^i)^\top \mu$. The noisy data distribution at time $t$ is then formulated based on the previous time step $t-1$ as follows:

\begin{equation}
p(\tilde{x}_t \mid \tilde{x}_{t-1}, \mathcal{P}) = \prod_{i=1}^{N_\mathcal{M}} \mathcal{N}(\tilde{x}_t^i; \tilde{x}_{t-1}^i, \lambda_t (\mathcal{H}^i)^\top \Sigma),
\end{equation}

\begin{equation}
    p(v_t \mid v_{t-1}, \mathcal{P}) = \prod_{i=1}^{N_\mathcal{M}} \mathcal{C}(v_t^i \mid (1 - \lambda_t) v_{t-1}^i + \lambda_t / K_a),
\end{equation}

\begin{equation}
    p(b_t \mid b_{t-1}, \mathcal{P}) = \prod_{i=1}^{N_\mathcal{M} \times N_\mathcal{M}} \mathcal{C}(b_t^{i} \mid (1 - \lambda_t) b_{t-1}^{i} + \lambda_t / K_b)
\end{equation}

where $K_a$ and $K_b$ represent the number of atom types and bond types, respectively. The perturbed structure is subsequently fed into the prediction model, and the reconstruction loss at time $t$ is derived from the KL divergence:

\[
L_t^{(x)} = \|x_0 - \hat{x}_0\|^2, \quad
L_t^{(v)} = \sum_{k=1}^{K_a} c(v_t, v_0)_k \log \frac{c(v_t, v_0)_k}{c(v_t, \hat{v}_0)_k},
\]
\[
L_t^{(b)} = \sum_{k=1}^{K_b} c(b_t, b_0)_k \log \frac{c(b_t, b_0)_k}{c(b_t, \hat{b}_0)_k},
\]

where $(x_0, v_0, b_0)$, $(x_t, v_t, b_t)$, and $(\hat{x}_0, \hat{v}_0, \hat{b}_0)$ denote the true atom positions, types, and bonds at times $0$, $t$, and predicted values, respectively. The categorical distribution coefficients $c$ depend on $\tilde{\alpha}_t$ and $1 - \tilde{\alpha}_t$. The overall loss is then defined as:

\[
L_t = L_t^{(x)} + \gamma_v L_t^{(v)} + \gamma_b L_t^{(b)},
\]

where $\gamma_v$ and $\gamma_b$ are the weights for the reconstruction losses of atom types and bond types, respectively.

\subsection{RAFT}
\label{raft}

\paragraph{Problem Setup}
We begin with a generative model $G_0 = g(w_0, x)$ parameterized by $w_0$, which takes an input $x \in \mathcal{X}$ and generates an output $y \in \mathcal{Y}$ according to the distribution $P_{G_0}^{1 / \lambda}(y \mid w_0, x)$, where $\lambda$ serves as a temperature parameter for controlling diversity. Additionally, we assume the existence of a reward function $r(x, y)$ that evaluates any input-output pair $(x, y)$.

To adapt the generative model, we utilize this reward function to guide $g(w, x)$. Denoting $p_g(y \mid w, x)$ as the conditional distribution for a given input $x$ under parameter $w$, and considering a training input distribution $\mathcal{D}$, the objective is expressed as:

\begin{equation}
\label{eq:alignment}
    \max_w \mathbb{E}_{x \sim \mathcal{D}, y \sim p_g(\cdot \mid w, x)} r(x, y).
\end{equation}

\paragraph{RAFT: Reward rAnked FineTuning}
We exploit the Reward rAnked FineTuning (RAFT) \cite{dong2023raft} approach, which combines sample ranking based on rewards with supervised fine-tuning (SFT). For simplicity, this method assumes that the generative model is capable of achieving the maximum reward for each input target protein $x$. Thus, each input $x \in \mathcal{X}$ can be handled independently, and the solution to Equation \ref{eq:alignment} is:

\begin{equation}
    p_g(\cdot \mid w^*, x) =
    \begin{cases}
        1 &\text{if } y = \arg\max_{y \in \mathcal{Y}} r(x, y), \\
        0 &\text{otherwise.}
    \end{cases}
\end{equation}

The RAFT learning process consists of three stages for each training iteration $t+1$:

\paragraph{Step 1: Data Collection.} 
We sample a batch of prompts $\mathcal{D}_t = \{x_1^t, \ldots, x_B^t\}$ from $\mathcal{X}$ and generate outputs $y_1, \ldots, y_K \sim P_{G_t}^{1 / \lambda}(\cdot \mid w_t, x_i^t)$ for each prompt $x_i^t \in \mathcal{D}_t$, where $\lambda$ controls output diversity.

\paragraph{Step 2: Data Ranking.}
Using the reward function, we compute the rewards $\{r(x, y_1), \ldots, r(x, y_K)\}$ for each $x \in \mathcal{D}_t$. The top-ranked output (generated ligand) $y := \arg\max_{y_j \in \{y_1, \ldots, y_K\}} r(x, y_j)$ is selected for each protein. We then construct a subset $\mathcal{B}$ of size $b$.

\paragraph{Step 3: Model Fine-Tuning.}
The model is fine-tuned on the subset $\mathcal{B}$, after which the next stage begins.

Implementing this method on the DecompDiff model \cite{guan2024decompdiff} has some challenges that we will discuss in the next section (\nameref{implementation}).

\subsection{Practical Implementation}
\label{implementation}

\begin{figure*}
    \centering
    \includegraphics[width=1\linewidth]{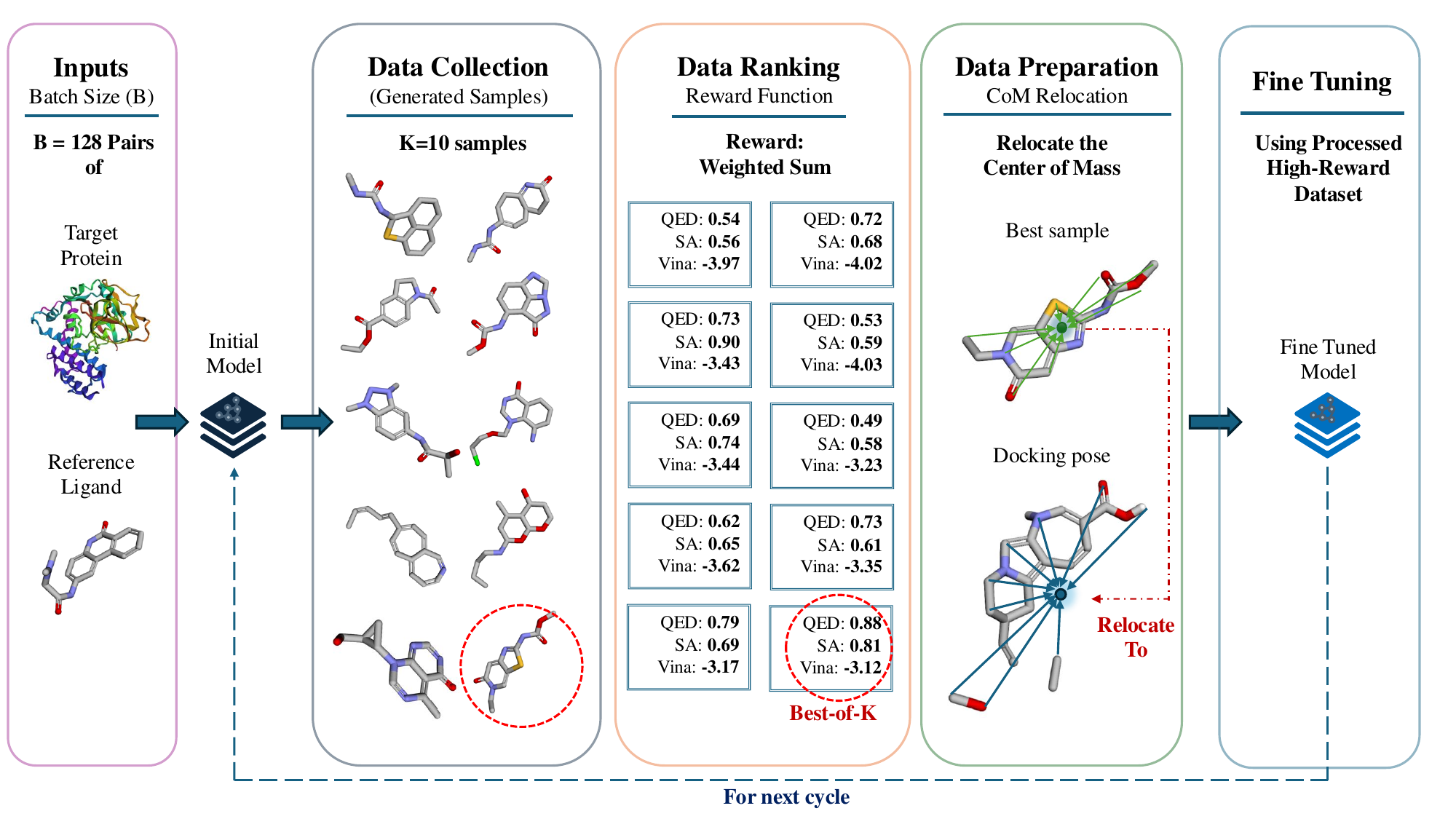}
    \caption{The BoKDiff framework: From left to right, (1) \textbf{Inputs}: Randomly select a batch of the desired size from the training set. (2) \textbf{Data Collection}: Generate $K$ samples for each input pair. (3) \textbf{Data Ranking}: Compute the desired metrics—approximations of QED, SA, and the Vina Docking score—for the generated samples. Rank them using a weighted sum approach; \textit{this example emphasizes QED and SA for the final ranking.} (4) \textbf{Data Preparation}: For the highest-reward sample, retrieve its docking pose and adjust its position by aligning its center of mass (CoM) to the CoM of the docking pose (See \nameref{preparation} section). Perform final preprocessing by extracting the corresponding sub-pockets and sub-structures for the high-reward pairs. (5) \textbf{Fine Tuning}: Update the model parameters using a small learning rate and limited epochs. This process is demonstrated for the target protein \textcolor{blue}{\textit{5ws0\_B\_rec}}.}
    \label{fig:bokdiff}
\end{figure*}

To implement the RAFT method on the DecompDiff model, we adhered to the three primary alignment steps while incorporating an additional data preparation stage prior to fine-tuning (See Figure \ref{fig:bokdiff}). To ensure a robust comparison and minimize deviations from the reference model, we leveraged the existing implementation for sampling and evaluation. The following sections provide a detailed breakdown of each step, highlighting our approach and key considerations throughout the process.

\subsubsection{Data Collection}

The original sampling scripts were designed for test set extraction, limited to 100 samples based on a provided indexer file. To enable training, we extended this approach by creating an analogous indexer file for the training set, containing 100K samples from the raw CrossDocked2020 dataset (see \nameref{dataset} section).

The sampling script was adapted to generate K=10 samples per training index. Invalid samples, which failed molecular reconstruction, were excluded from further processing.
    
\subsubsection{Data Ranking}

To evaluate generated molecules, we defined a reward function based on three metrics: drug-likeness (QED), synthesizability (SA), and docking score. These metrics were selected to optimize molecular properties while ensuring favorable binding affinity. As experimental validation of novel ligands is infeasible, we relied on established computational approximations consistent with prior studies \cite{guan2024decompdiff}.

QED and SA scores were computed using the \textit{RDKit} package and normalized to a [0, 1] range, with higher values indicating better performance. Docking scores, derived using the \textit{Vina} package, were normalized via min-max scaling across the K=10 generated samples per input. The scaled docking scores were inverted (lower values more favorable) for consistency in the reward calculation.

The final reward value was computed using a weighted sum of the metrics, with the weights treated as tunable hyperparameters during the alignment process. The highest-reward sample and its corresponding docking pose were selected for subsequent processing, serving as critical inputs for the later stages of our workflow.
    
\subsubsection{Data Preparation}
\label{preparation}

\paragraph{Challenges}
The model relies on detailed meta-information about protein-ligand interactions, such as protein pockets, ligand substructures, and fragment-specific details (e.g., atom types, coordinates, and bonds). However, a critical issue arises when generated ligands are not positioned close to the target protein in 3D space, making it difficult to identify protein binding pockets or extract ligand substructures. For instance, determining ligand “arms” depends on their proximity to the protein. Without proper positioning, these processes fail, compromising result reliability.

\paragraph{Initial Solution}
A preliminary solution involved replacing the original ligand structure with its docking pose generated by tools like Vina. Docking ensured that the ligand was placed near the protein. However, Vina often altered the ligand structure by modifying bonds, adding or removing atoms (e.g., hydrogen atoms), or reconstructing components. These changes frequently led to errors in substructure extraction or generated molecules that deviated significantly from the original design.

\paragraph{Proposed Modifications}
To address these challenges, we introduced two key adjustments:

\begin{enumerate}
    \item \textbf{Aligning the Ligand’s Center of Mass (CoM)}

    To preserve the ligand’s structure while ensuring proximity to the protein, the ligand’s CoM was aligned with the CoM of its docking pose. Non-heavy atoms were removed from the docking pose to calculate a cleaned CoM, which was then used to relocate the original ligand’s CoM near the protein binding site. This approach maintained the ligand’s integrity while achieving spatial alignment.

    \item \textbf{Adjusting Extraction Parameters}

    Strict parameters for identifying protein sub-pockets and ligand substructures, such as Cutoff and Radius, were dynamically adapted. Using a modified version of the Alphaspace2 package, we determined minimum values for these parameters to ensure successful extraction of the closest sub-components.
\end{enumerate}

An additional alternative was explored, wherein the known reference ligand from the dataset was used instead of the docking pose for evaluation. This approach provided a comparative baseline and is discussed in greater detail in the ablation studies (\nameref{ab:data_prep} section). 
    
\subsubsection{Model Fine-Tuning}

We initialized our approach using the best-trained checkpoint of the DecompDiff model. This checkpoint was first employed to generate samples and perform the preliminary steps described earlier. For fine-tuning, we constructed a high-reward dataset during the training process and reloaded the checkpoint with this curated data. Fine-tuning was conducted with a small learning rate and limited to 1,000 iterations to reduce the risk of overfitting while preserving the original model’s distributional characteristics.

In contrast to the original DecompDiff model’s training regime of 500,000 iterations, our fine-tuning process was deliberately constrained to maintain computational efficiency. Further details on the experimental setup and hyperparameter configurations are provided in the \nameref{ex_details} section.

\section{Experiments}
\label{results}

\subsection{Experimental Setup}
\subsubsection{Dataset}
\label{dataset}
Building on prior studies \cite{luo20213d, peng2022pocket2mol, guan2024decompdiff}, we utilized the filtered CrossDocked2020 dataset \cite{francoeur2020three} to fine-tune our base model and evaluate the performance of BoKDiff. The dataset, prepared according to the methodology outlined by Luo et al. \cite{luo20213d}, includes only complexes with high-quality docking poses (RMSD $<$1Å) and proteins with low sequence similarity (sequence identity $<$30\%). It comprises 100,000 high-quality complexes for training and 100 novel proteins for evaluation.

\subsubsection{Implementation details}
\label{ex_details}

All experiments were conducted on a single GeForce RTX 3090 GPU, with a batch size ($\mathcal{B}$) of 128. Generating 10 ligand samples per target protein required approximately 10 hours per batch. The data ranking step, including the calculation of Vina docking scores, took about 3 minutes per index (up to 10 samples), adding an additional 4 hours for a batch of 128 samples. Data preprocessing was integrated at the end of the ranking stage.

Model fine-tuning, performed for 1,000 epochs with a small learning rate of $1 \times 10^{-6}$, was computationally efficient due to the model’s size (approximately 5 million parameters), taking only 10 minutes. Final evaluation on the test dataset of 100 samples required about 4 hours, consistent with the data ranking runtime. To expedite ablation studies, evaluations were limited to 30 test samples.

\subsubsection{Baselines}
To evaluate the performance of the fine-tuned BoKDiff model, we compared it with several established generative frameworks. liGAN \cite{ragoza2022generating} uses a convolutional neural network (CNN)-based variational autoencoder to encode both ligands and receptors into a shared latent space, enabling the generation of atomic densities for ligands. Autoregressive models such as AR \cite{luo20213d}, Pocket2Mol \cite{peng2022pocket2mol}, and GraphBP \cite{liu2022generating} leverage graph neural networks (GNNs) to iteratively update atom embeddings. Diffusion-based models like TargetDiff \cite{guan20233d} and DecompDiff \cite{guan2024decompdiff} also rely on GNNs, with DecompDiff introducing decomposed priors to predict atom type, position, and bonding while maintaining structural validity through guided generation processes.

\subsubsection{Evaluation}
\label{evaluation}
We assess the quality of generated molecules by examining three key aspects: target binding affinity, molecular properties, and molecular conformation. Target binding affinity is evaluated using AutoDock Vina \cite{eberhardt2021autodock}, following protocols established in prior studies \cite{luo20213d, ragoza2022generating}. Three metrics are reported: Vina Score, which represents the direct binding affinity; Vina Min, reflecting affinity post-local structural minimization; Vina Dock, assessing affinity after re-docking the ligand.

For molecular properties, we analyze drug-likeness (QED) \cite{bickerton2012quantifying}, and synthetic accessibility (SA) \cite{ertl2009estimation}. A Success Rate metric is used to evaluate practical drug-design readiness, capturing the proportion of molecules meeting the thresholds: QED $> 0.25$, SA $> 0.59$, and Vina Dock $< \text{--} 8.18$, as defined by Jin et al. \cite{jin2020multi}. 

To assess molecular conformation, we compute the Jensen-Shannon divergence (JSD) between the atom/bond distance distributions of generated molecules and reference ligands, providing insight into the structural alignment of generated molecules with biologically relevant configurations. This comprehensive evaluation ensures the generated molecules are not only structurally valid but also possess favorable binding affinities and drug-like properties suitable for practical applications.

\subsection{Main Results}

We evaluate the performance of our model across Vina metrics and molecular properties. As shown in Table \ref{tab:main}, the aligned version of DecompDiff, termed BoKDiff, achieves notable improvements in the QED metric and several binding-related metrics. While our method did not enhance the SA property, the overall success rate of BoKDiff surpasses all baselines, including the original DecompDiff model. It is important to highlight that these results were obtained using our optimal weight configuration (QED: 1, SA: 0, Vina: 0). This demonstrates that aligning the model primarily with respect to the QED metric not only improves drug-likeness (QED) but also positively impacts Vina metrics and the final success rate.

\begin{table*}[ht]
\caption{Comparison of properties between reference molecules, molecules generated by our model, and baseline approaches. Larger values ($\uparrow$) or smaller values ($\downarrow$) indicate better performance, depending on the metric. The top two results for each property are highlighted using \textbf{bold} text (best) and \underline{underlined} text (second best). This experiment was conducted using our optimal weight configuration (QED: 1, SA: 0, Vina: 0).}
\label{tab:main}
\centering
\begin{tabular}{l|cc|cc|cc|cc|cc|c}
\hline
\multirow{2}{*}{Methods} & \multicolumn{2}{c|}{Vina Score ($\downarrow$)} & \multicolumn{2}{c|}{Vina Min ($\downarrow$)}   & \multicolumn{2}{c|}{Vina Dock ($\downarrow$)}  & \multicolumn{2}{c|}{QED ($\uparrow$)}      & \multicolumn{2}{c|}{SA ($\uparrow$)}       & Success Rate ($\uparrow$)    \\
                         & Mean           & Median         & Mean           & Median         & Mean           & Median         & Mean          & Median        & Mean          & Median        & Mean            \\ \hline
Reference                & -6.36          & -6.46          & -6.71          & -6.49          & -7.45          & -7.26          & 0.48          & 0.47          & 0.73          & 0.74          & 25.0\%          \\ \hline
liGAN                    & -              & -              & -              & -              & -6.33          & -6.20          & 0.39          & 0.39          & 0.59          & 0.57          & 3.9\%           \\
GraphBP                  & -              & -              & -              & -              & -4.80          & -4.70          & 0.43          & 0.45          & 0.49          & 0.48          & 0.1\%           \\
AR                       & {\ul -5.75}    & -5.64          & -6.18          & -5.88          & -6.75          & -6.62          & {\ul 0.51}    & {\ul 0.50}    & {\ul 0.63}    & {\ul 0.63}    & 7.1\%           \\
Pocket2Mol               & -5.14          & -4.70          & -6.42          & -5.82          & -7.15          & -6.79          & \textbf{0.56} & \textbf{0.57} & \textbf{0.74} & \textbf{0.75} & 24.4\%          \\
TargetDiff               & -5.47          & \textbf{-6.30} & -6.64          & -6.83          & -7.80          & -7.91          & 0.48          & 0.48          & 0.58          & 0.58          & 10.5\%          \\
DecompDiff               & -5.67          & {\ul -6.04}    & {\ul -7.04}    & {\ul -7.09}    & {\ul -8.39}    & \textbf{-8.43} & 0.45          & 0.43          & 0.61          & 0.60          & {\ul 24.5\%}    \\
BoKDiff(Ours)            & \textbf{-5.92} & -5.89          & \textbf{-7.50} & \textbf{-7.17} & \textbf{-8.58} & {\ul -8.41}    & 0.48          & 0.49          & 0.60          & 0.60          & \textbf{26.0\%} \\ \hline
\end{tabular}
\end{table*}

We further assess molecular conformations by comparing bond distance distributions between generated molecules and their corresponding reference empirical distributions, as presented in Table \ref{tab:bonds}. The aligned model exhibits strong potential for generating more stable molecular conformations, driven solely by alignment based on the drug-likeness (QED) metric. These results underscore the effectiveness of our alignment strategy in enhancing both molecular properties and conformational stability.

\begin{table*}[ht]
\caption{Jensen-Shannon divergence between bond distance distributions of reference molecules and generated molecules. Lower values indicate better performance. Symbols “-”, “=”, and “:” represent single, double, and aromatic bonds, respectively. The top two results are highlighted using \textbf{bold} text (best) and \underline{underlined} text (second best). This experiment was conducted using the optimal weight configuration (QED: 1, SA: 0, Vina: 0).}
\label{tab:bonds}
\centering
\begin{tabular}{cccccccc}
\hline
Bond & liGAN & GraphBP & AR    & Pocket2Mol & TargetDiff     & DecompDiff     & BoKDiff(Ours)  \\ \hline
C-C  & 0.601 & 0.368   & 0.609 & 0.496      & 0.369          & {\ul 0.359}    & \textbf{0.308} \\
C=C  & 0.665 & 0.530   & 0.620 & 0.561      & {\ul 0.505}    & 0.537          & \textbf{0.346} \\
C-N  & 0.634 & 0.456   & 0.474 & 0.416      & 0.363          & {\ul 0.344}    & \textbf{0.252} \\
C=N  & 0.749 & 0.693   & 0.635 & 0.629      & {\ul 0.550}    & 0.584          & \textbf{0.378} \\
C-O  & 0.656 & 0.467   & 0.492 & 0.454      & 0.421          & {\ul 0.376}    & \textbf{0.255} \\
C=O  & 0.661 & 0.471   & 0.558 & 0.516      & 0.461          & {\ul 0.374}    & \textbf{0.330} \\
C:C  & 0.497 & 0.407   & 0.451 & 0.416      & {\ul 0.263}    & \textbf{0.251} & 0.297          \\
C:N  & 0.638 & 0.689   & 0.552 & 0.487      & \textbf{0.235} & 0.269          & {\ul 0.237}    \\ \hline
\end{tabular}
\end{table*}

\subsection{Best-of-N}
Instead of curating a high-reward dataset by selecting the best-of-K samples and fine-tuning the model based on this dataset, an alternative approach skips the fine-tuning step entirely by directly selecting the best sample from $N$  generated samples. It is important to differentiate between these two approaches: the Best-of-K strategy involves selecting the best sample to fine-tune the model, while the Best-of-N strategy selects the best sample without any subsequent fine-tuning. In the Best-of-N approach, the reward model plays a critical role as the selection of the best sample relies entirely on its reward value.

Previous studies, including those in the BoNBoN paper, underscore the effectiveness of Best-of-N (BoN) sampling \cite{gui2024bonbon}. Notably, BoN sampling has been shown to approximate the sampling distribution of the policy learned through reinforcement learning from human feedback (RLHF) \cite{ouyang2022training}, demonstrating both theoretical soundness and practical efficiency.

\begin{figure}
    \centering
    \includegraphics[width=1\linewidth]{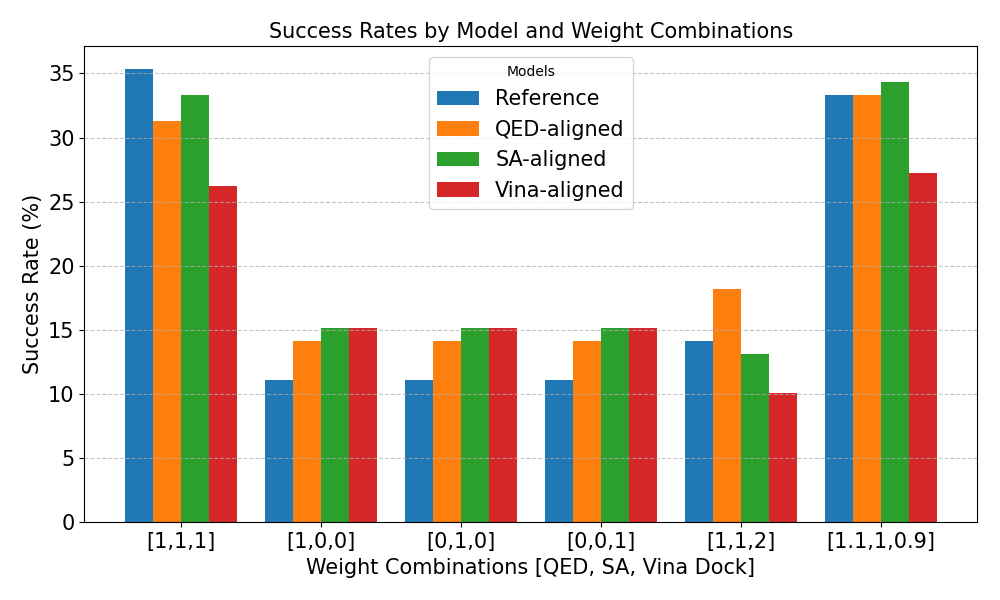}
    \caption{Success rates of reference and aligned models using the best-of-N strategy under various weight combinations for reward value determination. For all models,  $N = 20$.}
    \label{fig:bon}
\end{figure}

Our results in Table \ref{tab:bon} validate the effectiveness of the BoN sampling approach, showing that the model performance stabilizes with $N = 30$. However, $N = 20$ provides a close approximation while significantly reducing computational costs, making it a practical stopping point. Notably, the Vina values in the BoN table (Table \ref{tab:bon}) are worse compared to those in the BoK-aligned table (Table \ref{tab:main}). This discrepancy arises because, in the BoN approach, the mean and median are calculated based only on the best samples, which are fewer in number compared to the BoK-aligned method. In contrast, Table \ref{tab:main} considers all 10 generated samples per protein when calculating evaluation metrics. While considering all samples results in better Vina values; the QED, SA, and success rate values are significantly better in the BoN approach. Furthermore, Figure \ref{fig:bon} compares the performance of the reference model and aligned models, illustrating that while BoN sampling offers substantial advantages, fine-tuned models retain unique benefits in specific scenarios, highlighting the complementary nature of these strategies.

\begin{table*}[ht]
\caption{Comparison of Best-of-N performance on the reference model for various values of $N$ . Results are provided for the weight combination prioritizing QED, while other combinations emphasizing SA or Vina Dock are omitted due to their similarity to [1, 0, 0]. The optimal weight set, [1.1, 1, 0.9], was identified through a grid search centered around the baseline [1, 1, 1]. Weight set legend: [QED, SA, Vina Dock].}
\label{tab:bon}
\centering
\resizebox{\textwidth}{!}{
\begin{tabular}{c|c|cc|cc|cc|cc|cc|c}
\hline
\multirow{2}{*}{\begin{tabular}[c]{@{}c@{}}N \\ (Best-of-N)\end{tabular}} & \multirow{2}{*}{Weights} & \multicolumn{2}{c|}{Vina Score ($\downarrow$)} & \multicolumn{2}{c|}{Vina Min ($\downarrow$)}   & \multicolumn{2}{c|}{Vina Dock ($\downarrow$)}  & \multicolumn{2}{c|}{QED ($\uparrow$)}      & \multicolumn{2}{c|}{SA ($\uparrow$)}       & Success Rate ($\uparrow$)     \\
                                                                          &                          & Mean           & Median         & Mean           & Median         & Mean           & Median         & Mean          & Median        & Mean          & Median        & Mean             \\ \hline
10                                                                        & {[}1, 1, 1{]}            & -5.57          & -5.42          & -6.38          & -6.22          & -7.7           & -7.71          & 0.61          & 0.63          & 0.71          & 0.67          & 22\%             \\
15                                                                        & {[}1, 1, 1{]}            & \textbf{-5.82} & \textbf{-5.68} & -6.64          & \textbf{-6.54} & -7.9           & -7.77          & 0.63          & 0.65          & 0.73          & 0.74          & 32\%             \\
20                                                                        & {[}1, 1, 1{]}            & -5.64          & -5.67          & -6.54          & -6.42          & -7.95          & -7.91          & 0.65          & 0.66          & 0.75          & 0.76          & 35.35\%          \\
25                                                                        & {[}1, 1, 1{]}            & -5.74          & -5.58          & \textbf{-6.67} & -6.36          & \textbf{-8.09} & -7.8           & 0.66          & 0.7           & \textbf{0.76} & \textbf{0.78} & 34.69\%          \\
30                                                                        & {[}1, 1, 1{]}            & -5.63          & -5.52          & -6.65          & -6.39          & -8.03          & -7.77          & 0.67          & 0.72          & \textbf{0.76} & \textbf{0.78} & 35.71\%          \\
10                                                                        & {[}1, 0, 0{]}            & -5.22          & -5.12          & -6.08          & -5.93          & -7.17          & -7.18          & 0.48          & 0.48          & 0.64          & 0.64          & 14\%             \\
15                                                                        & {[}1, 0, 0{]}            & -5.19          & -5.05          & -6.05          & -5.95          & -7.16          & -7.15          & 0.48          & 0.47          & 0.64          & 0.64          & 11\%             \\
20                                                                        & {[}1, 0, 0{]}            & -5.13          & -5.06          & -6.01          & -5.88          & -6.97          & -7.09          & 0.48          & 0.48          & 0.65          & 0.64          & 11.11\%          \\
25                                                                        & {[}1, 0, 0{]}            & -5.18          & -5.05          & -6.02          & -5.96          & -7.16          & -7.22          & 0.48          & 0.48          & 0.65          & 0.64          & 12.24\%          \\
30                                                                        & {[}1, 0, 0{]}            & -5.18          & -5.12          & -6.01          & -6.03          & -7.23          & -7.33          & 0.51          & 0.51          & 0.65          & 0.64          & 12.24\%          \\ \hline
10                                                                        & {[}1.1, 1, 0.9{]}        & -5.56          & -5.42          & -6.34          & -6.22          & -7.61          & -7.58          & 0.61          & 0.63          & 0.71          & 0.68          & 23\%             \\
15                                                                        & {[}1.1, 1, 0.9{]}        & -5.8           & -5.58          & -6.62          & \textbf{-6.54} & -7.83          & -7.7           & 0.64          & 0.65          & 0.73          & 0.74          & 30\%             \\
20                                                                        & {[}1.1, 1, 0.9{]}        & -5.69          & -5.66          & -6.57          & -6.42          & -7.97          & -7.91          & 0.65          & 0.66          & 0.74          & 0.76          & 33.33\%          \\
25                                                                        & {[}1.1, 1, 0.9{]}        & -5.73          & -5.58          & -6.65          & -6.36          & -8.06          & -7.8           & 0.66          & 0.71          & \textbf{0.76} & \textbf{0.78} & 33.67\%          \\
30                                                                        & {[}1.1, 1, 0.9{]}        & -5.61          & -5.52          & \textbf{-6.67} & -6.43          & -8.06          & \textbf{-7.87} & \textbf{0.67} & \textbf{0.73} & \textbf{0.76} & \textbf{0.78} & \textbf{37.76\%} \\ \hline
\end{tabular}}
\end{table*}

\subsection{Ablation Studies}
We explore several hyperparameters critical to the performance of our BoKDiff framework. These include weight configurations used to compute the final reward value for ranking and selecting the top generated samples, batch size ($B$), which impacts the fine-tuned model by determining the size of the high-reward dataset utilized as input, and the number of iterations for repeating the framework’s entire process. Additionally, we investigate different approaches to resolving the proximity issue by relocating the center of mass. In the following sections, we present ablation studies conducted on 30 test set samples, focusing specifically on the three core metrics underlying our method: QED, SA, and Vina Docking Score.

\subsubsection{Weight Configurations}
We conducted experiments using five distinct weight configurations to calculate the final reward value. Tables \ref{tab:weights_iter1} and \ref{tab:weights_iter2} summarize the results for iterations one and two, respectively. These experiments reveal an interesting trend: focusing exclusively on molecular property metrics, such as QED or SA, often leads to concurrent improvements in other metrics. Notably, the best-performing weight configuration was (QED:1, SA:0, Vina Dock:0). This result highlights that prioritizing drug-likeness (QED) can also positively influence synthesizability (SA) and docking affinity, demonstrating a beneficial correlation between these objectives.

\begin{table*}[ht]
\caption{Comparison of the impact of different reward weight configurations on final results. These results correspond to the \textbf{first iteration}, using 30 data samples selected from the test set. (Accumulated columns represent the simple summation (with equal weights) of three metrics, where the Vina Dock scores are normalized by dividing them by -10 to ensure they fall within the [0, 1] range.)}
\label{tab:weights_iter1}
\centering
\begin{tabular}{l|cc|cc|cc|cc}
\hline
\multirow{2}{*}{\begin{tabular}[c]{@{}l@{}}Weights\\ (QED, SA, Vina Dock)\end{tabular}} & \multicolumn{2}{c|}{Vina ($\downarrow$)} & \multicolumn{2}{c|}{QED ($\uparrow$)} & \multicolumn{2}{c|}{SA ($\uparrow$)} & \multicolumn{2}{c}{Accumulated ($\uparrow$)} \\
                                                                                        & Mean        & Median      & Mean       & Median      & Mean       & Median     & Mean           & Median         \\ \hline
(1, 1, 1)                                                                               & -7.048      & -7.338      & \underline{0.503}      & \underline{0.512}       & \textbf{0.659}      & 0.64       & 1.867          & 1.886          \\
(1, 0, 0)                                                                               & \underline{-7.135}      & \textbf{-7.649}      & \textbf{0.509}      & \textbf{0.518}       & \underline{0.657}      & \textbf{0.65}       & \textbf{1.880}          & \textbf{1.933}          \\
(0, 1, 0)                                                                               & \textbf{-7.212}      & -7.524      & 0.494      & 0.497       & 0.656      & \underline{0.645}      & \underline{1.871}          & \underline{1.894}          \\
(0, 0, 1)                                                                               & -7.018      & -7.532      & 0.494      & 0.488       & 0.653      & 0.64       & 1.849          & 1.881          \\
(1, 1, 2)                                                                               & -7.197      & \underline{-7.559}      & 0.497      & 0.495       & 0.653      & 0.64       & 1.870          & 1.891          \\ \hline
\end{tabular}
\end{table*}

\begin{table*}[ht]
\caption{Comparison of the impact of different reward weight configurations on final results. These results correspond to the \textbf{second iteration}, using 30 data samples selected from the test set. (Accumulated columns represent the simple summation (with equal weights) of three metrics, where the Vina Dock scores are normalized by dividing them by -10 to ensure they fall within the [0, 1] range.)}
\label{tab:weights_iter2}
\centering
\begin{tabular}{l|cc|cc|cc|cc}
\hline
\multirow{2}{*}{\begin{tabular}[c]{@{}l@{}}Weights\\ (QED, SA, Vina Dock)\end{tabular}} & \multicolumn{2}{c|}{Vina ($\downarrow$)} & \multicolumn{2}{c|}{QED ($\uparrow$)} & \multicolumn{2}{c|}{SA ($\uparrow$)} & \multicolumn{2}{c}{Accumulated ($\uparrow$)} \\
                                                                                        & Mean        & Median      & Mean       & Median      & Mean       & Median     & Mean           & Median         \\ \hline
(1, 1, 1)                                                                               & -7.1        & -7.328      & 0.502      & 0.517       & 0.645      & \textbf{0.64}       & 1.857          & 1.890          \\
(1, 0, 0)                                                                               & -7.108      & \underline{-7.561}      & \underline{0.511}      & \textbf{0.526}       & \textbf{0.648}      & 0.63       & \textbf{1.870}          & \underline{1.912}          \\
(0, 1, 0)                                                                               & \underline{-7.111}      & \textbf{-7.607}      & \textbf{0.515}      & \underline{0.525}       & 0.641      & 0.63       & \underline{1.867}          & \textbf{1.916}          \\
(0, 0, 1)                                                                               & \textbf{-7.139}      & -7.476      & 0.498      & 0.511       & 0.644      & 0.62       & 1.856          & 1.879          \\
(1, 1, 2)                                                                               & -7.024      & -7.165      & 0.502      & 0.515       & \underline{0.646}      & 0.63       & 1.850          & 1.862          \\ \hline
\end{tabular}
\end{table*}

\subsubsection{Batch Size}
We analyzed the impact of different batch sizes on performance using the optimal weight configuration (QED:1, SA:0, Vina Dock:0). Testing batch sizes of 128 and 256 revealed that while larger batch sizes may offer marginal improvements, the gains were inconsistent and, in some cases, led to overfitting, reducing performance on certain metrics. To address this, incorporating a KL divergence term into the reward function could help stabilize the model and prevent distribution drift, as discussed in \nameref{limitation} section.


\subsubsection{Iteration Number}
\label{ab:iter}
Experiments conducted over three consecutive iterations reveal that running the proposed framework multiple times can lead to incremental improvements in some metrics. However, the most significant progress is typically observed in the first iteration, making it a reasonable stopping point for evaluating the test set and reporting results.

The fixed atom number constraint in the reference model (DecompDiff) and its tendency to favor simpler molecules, particularly with less complex arms, limit its effectiveness for further iterations. These constraints suggest that the model may not fully exploit the iterative process to generate increasingly optimized molecules.


\subsubsection{Data Preparation Method}
\label{ab:data_prep}
To address the proximity issue between the generated ligand and the target protein, we tested two innovative approaches: one based on relocating the ligand according to the center of mass (CoM) of the docking pose, and another using the reference ligand provided in the dataset. As shown in Table \ref{tab:versions}, the docking pose method demonstrated greater potential for improving performance. Since the docking pose is inherently an approximation influenced by random seeds, this approach can be further extended to accommodate multiple poses generated under varying conditions. For additional details, see Limitations and Future Work (\nameref{limitation} section).

\begin{table*}[ht]
\caption{Comparison of cumulative results for different methods of relocating the Center of Mass (CoM), based on the Docking Pose CoM and the Reference Ligand CoM. (Accumulated columns represent the simple summation (with equal weights) of three metrics, where the Vina Dock scores are normalized by dividing them by -10 to ensure they fall within the [0, 1] range.)}
\label{tab:versions}
\centering
\begin{tabular}{l|c|c|cc|cc|cc|cc}
\hline
\multicolumn{1}{c|}{}                          &                             &                                                                                   & \multicolumn{2}{c|}{Vina ($\downarrow$)} & \multicolumn{2}{c|}{QED ($\uparrow$)} & \multicolumn{2}{c|}{SA ($\uparrow$)} & \multicolumn{2}{c}{Accumulated ($\uparrow$)}        \\
\multicolumn{1}{c|}{\multirow{-2}{*}{Method}} & \multirow{-2}{*}{Iteration} & \multirow{-2}{*}{\begin{tabular}[c]{@{}c@{}}(QED, SA,\\  Vina Dock)\end{tabular}} & Mean        & Median      & Mean       & Median      & Mean       & Median     & Mean                          & Median \\ \hline
Docking pose                                   & 1                           & (1, 1, 2)                                                                         & \textbf{-7.197}      & \underline{-7.559}      & \underline{0.497}      & 0.495       & \textbf{0.653}      & \textbf{0.64}       & \textbf{1.870} & \underline{1.891}  \\
Reference ligand                               & 1                           & (1, 1, 2)                                                                         & -7.07       & \textbf{-7.8}        & 0.49       & 0.481       & \textbf{0.653}      & \textbf{0.64}       & 1.850 & \textbf{1.901}  \\
Docking pose                                   & 2                           & (1, 1, 2)                                                                         & \underline{-7.024}      & -7.165      & \textbf{0.502}      & \textbf{0.515}       & \underline{0.646}      & 0.63       & 1.850 & 1.862  \\
Reference ligand                               & 2                           & (1, 1, 2)                                                                         & -6.76       & -7.343      & 0.485      & \underline{0.498}       & 0.634      & 0.63       & 1.795 & 1.862  \\ \hline
\end{tabular}
\end{table*}

\section{Limitation and Future works}
\label{limitation}
The proposed framework has three notable limitations and areas for potential improvement:

\begin{enumerate}
    \item \textbf{Incorporating KL Divergence in the Reward Function:} Introducing a KL divergence term in the reward function could help mitigate distribution drift from the reference model, ensuring more stable improvements with each iteration. As observed in the ablation study ( \nameref{ab:iter} section), even the prioritized metric in the reward function does not consistently improve across iterations. Adding this term may address the observed instability.
    \item \textbf{Limitations of the DecompDiff Model:} The fixed atom number constraint and the model’s tendency to favor simpler molecular structures, such as those with less complex arms, restrict DecompDiff’s capacity to fully utilize the iterative optimization process (see \nameref{ab:iter} section). This limitation suggests that the model is not well-suited for generating highly intricate molecules, thereby constraining the overall performance of the proposed framework.
    \item \textbf{Enhanced Ligand Relocation and Transformation:} The current approach relocates generated ligands based on their center of mass, but a more robust solution would involve rotating ligand arms and reforming structures to align more accurately with the docking pose. This could be extended to include variations in binding sites, such as alternative docking poses or reference ligands (see \nameref{ab:data_prep} section). However, existing algorithms, such as those in the RDKit package, struggle with significant molecular shape differences between the original ligands and docking poses. Developing a novel algorithm to handle these transformations is a promising direction for future work.
    \item \textbf{Expanding the Reward Function and Optimization Methodology:} The current framework focuses solely on Vina Docking scores while excluding other relevant Vina metrics from the reward function. Furthermore, the traditional weighted sum approach for combining objectives could be replaced by multi-level optimization techniques. These advanced methods would allow for a more nuanced consideration of each objective, offering a more comprehensive approach to multi-objective optimization.
\end{enumerate}

Addressing these limitations could significantly enhance the performance, robustness, and practical applicability of the framework in drug discovery tasks.

\section{Conclusion}
In this work, we presented BoKDiff, a novel multi-objective optimization framework for aligning diffusion models to improve molecular generation for drug discovery. By leveraging a Best-of-K methodology, we focused on enhancing key molecular properties, including drug-likeness (QED), synthetic accessibility (SA), and docking affinity (Vina score). Our framework incorporates a refined data preparation method to address the proximity issue between generated ligands and target proteins, utilizing docking pose alignment and center-of-mass relocation. Through extensive experiments, we demonstrated that BoKDiff consistently outperforms baseline approaches, achieving high metrics across various configurations.

In addition to BoKDiff, we explored the Best-of-N (BoN) strategy, which offers a simpler implementation but involves higher computational costs during inference due to evaluating a larger set of generated samples. Despite this trade-off, BoN achieved superior results in both molecular property optimization and success rates compared to BoKDiff, highlighting its potential as a more robust method for challenging molecular design tasks. The effectiveness of BoN underscores the importance of exploring scalable sampling strategies for future advancements in generative molecular design.

BoKDiff, alongside the BoN strategy, paves the way for incorporating more flexible and powerful diffusion models, exploring advanced reward alignment strategies, and refining data preparation techniques to accommodate diverse docking poses. This work serves as a step toward bridging generative molecular design with practical drug discovery requirements, offering a promising framework for future advancements.




\bibliographystyle{plain}
\bibliography{bibliography}


\end{document}